\documentclass[aps,letterpaper,showpacs,showkeys,prc,twocolumn,nofootinbib,10pt]{revtex4-1}

\usepackage{graphicx}
\usepackage{amsmath}
\usepackage{amssymb}
\usepackage[sort&compress]{natbib}
\usepackage{ifthen}
\usepackage{hyperref} 
\usepackage{breakurl}
\usepackage{color}
\usepackage{horowitzarXiv}

\definecolor{darkgreen}{rgb}{0,.7,0}
\definecolor{linkblue}{rgb}{0.,0.,0.9333}

\hypersetup{
    pdffitwindow=true,      
    pdftitle={The Surprising Transparency of the sQGP at LHC},    
    pdfauthor={W. A. Horowitz and Miklos Gyulassy},     
    pdfnewwindow=true,      
    bookmarksnumbered=true,
    colorlinks=true,       
    linkcolor=red,          
    citecolor=darkgreen,        
    filecolor=magenta,      
    urlcolor=linkblue,           
    menucolor=blue,
}

\begin{document}

\title{The Surprising Transparency of the sQGP at LHC}

\date{\today}

\author{W.\ A.\ Horowitz}
\email{wa.horowitz@uct.ac.za}
\affiliation{Department of Physics, University of Cape Town, Private Bag X3, Rondebosch 7701, South Africa}
$\phantom{\email{}}$

\author{Miklos Gyulassy}
\affiliation{Department of Physics, Columbia University,\\538 West 120$^{th}$ Street, New York, NY 10027, USA}

\begin{abstract}
 We present parameter-free predictions of 
the nuclear modification factor, $R_{AA}^\pi(p_T,s)$, of high
 $p_T$ pions produced in Pb+Pb collisions at $\sqrt{s}_{NN}=2.76 \;{\rm
   and}\; 5.5$ ATeV based on the WHDG/DGLV
 (radiative+elastic+geometric fluctuation) jet energy loss model. 
The initial quark gluon plasma (QGP) density at LHC is
 constrained from a rigorous statistical analysis of PHENIX/RHIC $\pi^0 $ 
 quenching data at $\sqrt{s}_{NN}=0.2$ ATeV and the 
charged particle multiplicity at ALICE/LHC at 2.76 ATeV.
Our 
perturbative QCD tomographic theory predicts
 significant differences between jet quenching at RHIC and LHC energies, which are qualitatively consistent with the $p_T$-dependence and normalization---within the large systematic uncertainty---of the first {\em charged} hadron nuclear modification
 factor, $R^{ch}_{AA}$, data measured by ALICE.  
However, our constrained prediction of
 the central to peripheral pion modification, $R^\pi_{cp}(p_T)$, 
for which large systematic uncertainties associated with 
unmeasured  p+p reference data cancel, is found to be over-quenched 
relative to the charged hadron ALICE $R^{ch}_{cp}$ data in  the range $5<p_T<20$ GeV/c. The discrepancy challenges 
the two most basic jet tomographic assumptions:
(1) that the energy loss scales linearly with the initial local
 comoving QGP density, $\rho_0$, and (2) 
 that $\rho_0 \propto dN^{ch}(s,{\cal C})/dy$ 
is proportional to the observed global charged
 particle multiplicity per unit rapidity 
as a function of $\sqrt{s}$ and centrality class, ${\cal C}$.  
 Future LHC identified $(h=\pi,K,p)$
 hadron  $R^h_{AA}$ data (together with precise p+p,
 p+Pb, and $Z$ boson and direct photon Pb+Pb control data) 
are needed
to assess if the QGP produced at LHC is indeed less opaque to jets
than predicted by constrained extrapolations from RHIC.
\\[2ex]

\end{abstract}

\pacs{12.38.Mh, 24.85.+p, 25.75.-q}
\keywords{QCD, Relativistic heavy-ion collisions, Quark gluon plasma, Jet quenching, Jet Tomography}

\maketitle

\section{Introduction}
The first LHC  
Pb+Pb data at $\sqrt{s}=2.76$ ATeV \cite{Aamodt:2010jd,Aamodt:2010pb,Aamodt:2010pa,Collaboration:2010cz,Aad:2010bu,Chatrchyan:2011sx} provide 
important new consistency tests  
of dynamical models developed over the past two decades 
to predict multiparticle
observables in nuclear reactions at  $\sqrt{s}=0.2$ AGeV from 
the Relativistic Heavy Ion Collider (RHIC/BNL) \cite{Adcox:2004mh,Adams:2005dq,Back:2004je,Arsene:2004fa}. The new LHC energy frontier with Pb+Pb probes the physics
 of strongly coupled Quark Gluon Plasma (sQGP) \cite{Gyulassy:2004zy,Shuryak:2004cy,Muller:2006ee} 
at  densities approximately twice as high as at RHIC,  
with temperatures up to $T\sim 500$ MeV,  
well above the deconfinement transition region predicted by lattice
QCD \cite{Borsanyi:2010cj,Kaczmarek:2011zz}.

The theoretical understanding of the dynamical properties of this new 
form of matter is however far from complete.  
On the one hand, bulk radial and differential azimuthal anisotropic 
elliptic flow data of
low transverse momenta ($p\sim 3T_0\approx 1$ GeV)  partons
were found to be consistent with near ``perfect fluid'' flow and suggest 
highly nonperturbative physics of this new form of matter \cite{Luzum:2008cw,Schenke:2010rr}. 
This has led to 
proposals that the bulk properties of the sQGP may be better
approximated via strong coupling (supergravity dual) holography
models \cite{Gubser:1996de,Gubser:1998bc,Witten:1998zw,Kovtun:2004de,Gubser:2006bz,Herzog:2006gh,Herzog:2006se,Blaizot:2005fd,Gubser:2008as,Chesler:2008uy, Gubser:2009fc,Chesler:2010bi} 
than via perturbative QCD (pQCD) based quark and gluon quasiparticle 
Hard Thermal Loop (HTL) \cite{Braaten:1989mz,Blaizot:2000fc} approximations.  

On the other hand, 
short wavelength ($p_T \sim 10-20$ GeV/c) properties
of the sQGP, as measured via the nuclear modification 
factor of pions \cite{Adler:2003au,Adams:2003kv,Adare:2008cg},
were found to be well predicted by pQCD based HTL partonic 
radiative energy 
loss theory \cite{MGMP:1990ye,Wang:1991hta,Wang:1991xy,MGXW:1993hr,Zakharov:1997uu,Zakharov:1998sv,Wang:1998ww,Gyulassy:2000fs,Gyulassy:2000er,Wiedemann:2000za,Baier:2001yt,Vitev:2002pf,Djordjevic:2003zk,Djordjevic:2003be,Djordjevic:2005nh,Wang:2002ri,Armesto:2003jh,Majumder:2004pt,Wicks:2005gt,Qin:2007rn,Armesto:2009zi,Majumder:2009zu,Chen:2010te}. 
The nuclear modification factor of high transverse momentum 
hadron, $h$, fragments in $A+B\rightarrow h+X$ and centrality class ${\cal C}$
used to  probe the short wavelength dynamics in an sQGP is defined as
\begin{equation}
 R_{AB}^h(y,\vec{p}_T; \sqrt{s},{\cal C}) = 
\frac{dN^{A+B\rightarrow h}(y,\vec{p}_T,\sqrt{s},{\cal C})/dyd^2\vec{p}_T}{
T_{AB}({\cal C})d\sigma^{p+p\rightarrow h}(\sqrt{s})/dyd^2\vec{p}_T} \; .
\end{equation}
For a fixed  $\sqrt{s}$ center of mass (cm) energy (per nucleon pair) and
nucleon-nucleon (NN) inelastic cross section $\sigma_{NN}^{in}(\sqrt{s})$ the mean number of elementary binary NN collisions in centrality class ${\cal C}$ is given by 
$\sigma_{NN}^{in}T_{AA}$, where
\cite{Miller:2007ri} 
\begin{eqnarray}
T_{AB}({\cal C})&=& 
\left\langle \int {d^2\vec{x}_\perp}
T_A(\vec{x}_\perp -\frac{\vec{b}}{2}) T_B(\vec{x}_\perp +\frac{\vec{b}}{2})
\right\rangle_{b\in \cal C}
\end{eqnarray}
in terms of the Glauber nuclear thickness profile
$T_A(\vec{x}_\perp)=\int dz \rho_{A}(z,\vec{x}_\perp)$
and Wood-Saxon nuclear density $\rho_A$ normalized to $A$.

\begin{figure}[!hbp]
\includegraphics[width=\columnwidth]{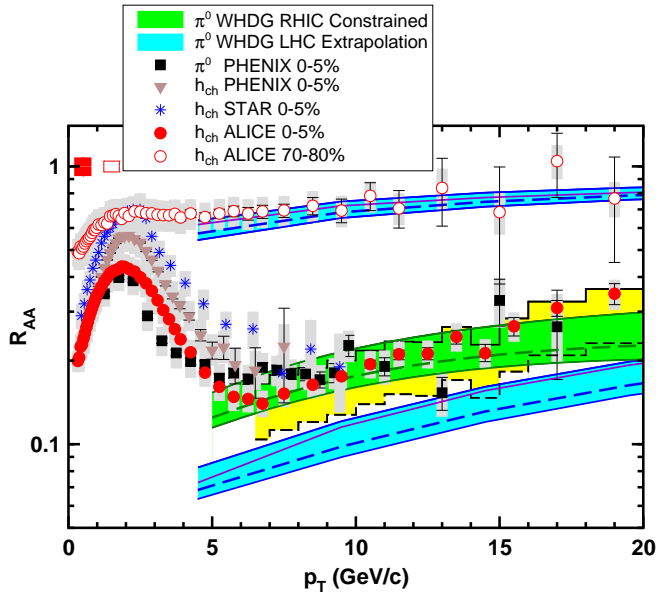}
\caption{\protect\label{fig:RAAnow}WHDG model \protect\cite{Wicks:2005gt} predictions (blue bands extrapolated from the RHIC constrained green band) 
for the nuclear modification factor of $\pi^0$ in Pb+Pb 2.76 ATeV LHC are compared to ALICE/LHC \protect\cite{Aamodt:2010jd} {\em charged} hadron nuclear modification data in central (red solid)
 and peripheral (open red)  reactions. 
The PHENIX/RHIC Au+Au$\rightarrow\pi^0$ nuclear modification
data \protect\cite{Adare:2008cg} are shown by black dots. 
The brown triangles and blue stars represent the charged hadron PHENIX \protect\cite{Adler:2003au} and STAR \protect\cite{Adams:2003kv} data, respectively. 
The blue band of WHDG predictions corresponds to the 1-$\sigma$ medium constraint set by PHENIX \protect\cite{Adare:2008cg} extrapolated to LHC via the ALICE charged particle rapidity density \protect\cite{Aamodt:2010pb}.  
The wide  
 yellow band is the current systematic error band of the (red dot) LHC data
due to the unmeasured p+p reference denominator.}
\end{figure}

In the absence of both initial state 
and final state nuclear 
interactions  $R_{AB}=1$. For
$p_T$ below some characteristic medium dependent transverse momentum
``saturation'' 
scale, $Q_s(p_T,\sqrt{s},A)$, 
the initial nuclear partonic distributions 
functions (PDFs) \cite{deFlorian:2003qf,Hirai:2007sx,Eskola:2009uj} $f_{a/A}(x=2 p_T/\sqrt{s},Q^2\sim p_T^2)< A f_{a/N}(x,Q^2)$  
are expected 
to be shadowed, leading to $R_{AA}<1$
because the incident flux of partons is less than $A$ times the free 
nucleon parton flux. Color Glass Condensate (CGC) models \cite{Blaizot:1987nc,McLerran:1993ni,Iancu:2003xm,Gyulassy:2004zy,Weigert:2005us,JalilianMarian:2005jf,McLerran:2008pe,Albacete:2010bs}
have been developed to predict $Q_s(p_T,\sqrt{s},A)$ related initial state
effects from first principles.  
While the magnitude of
$Q_s$ at LHC is uncertain and will
require future dedicated p+Pb control measurements to 
map out, current expectations are that  $Q_s<5$ GeV at LHC 
in the central rapidity region.
This should leave a  wide
jet tomographic kinematic window 
$10 < p_T< 200$ GeV in which nuclear modification 
should be dominated by final state parton energy loss and broadening
effects.  In this paper, we therefore assume that initial state nuclear effects 
can be neglected in the $10 < p_T< 20$ (i.e.\ $x>0.01$) range
explored by the first ALICE data \cite{Aamodt:2010jd}.
We note that from 
\fig{fig:RAAnow}, and as discussed in detail below, our RHIC constrained jet quenching
due to  
final state interactions alone already tends to 
over-predict the pion quenching at LHC 
and therefore 
leaves  no room for large additional shadowing/saturation
effects in the \cite{Albacete:2010bs,Deng:2010mv,Levai:2011qm} in this  $Q^2>100$ GeV$^2$ kinematic window---unless
the sQGP is much more transparent at LHC than expected from most extrapolations
of jet quenching phenomena from SPS and RHIC to LHC energies.

The main challenge to 
pQCD multiple collision theory of jet tomography 
and AdS/CFT 
jet holography
is how to construct 
a consistent approximate framework that can  
account simultaneously for the beam energy dependence from SPS to LHC energy
and for the nuclear system size, momentum, 
and centrality
dependence from $p+p$ to $U+U$ of four major classes of 
hard probe observables: 
(1) the light quark and gluon leading jet
quenching pattern as a function of the resolution scale $p_T$, (2) the
heavy quark flavor dependence of jet flavor tagged observables, and (3) the azimuthal dependence of high 
$p_T$ particles relative to the bulk reaction plane determined 
from \lowpt elliptic flow and higher azimuthal flow moments,
$v_n(p_T)$, and (4) corresponding di-jet observables. 

The first LHC heavy ion data on high transverse momentum spectra 
provide an important  milestone because they test for the first time the 
density or opacity dependence of light
quark and gluon jet quenching theory 
in a parton density range approximately twice as large
as that studied at RHIC. 
The 
 surprise from LHC is the relatively small difference observed
between  the RHIC \cite{Adler:2003au,Adams:2003kv,Adare:2008cg} and ALICE \cite{Aamodt:2010jd} LHC data
 on $R_{AA}(10<p_T<20\;{\rm GeV})$, as shown in \fig{fig:RAAnow}. In addition,
there is little difference from RHIC to LHC between the differential elliptic flow probe, $v_2(p_T< 2)$, 
as reported in
\cite{Aamodt:2010pa}.  
The rather striking similarities between
bulk and hard observables at RHIC and LHC pose significant 
 consistency challenges for both initial state production 
and dynamical modeling of the sQGP phase of matter.

In this paper, we focus on the 
puzzle posed by the similarity of inclusive light quark/gluon jet 
quenching at RHIC and LHC 
by performing a constrained extrapolation from RHIC using 
 the  WHDG model \cite{Wicks:2005gt} to predict $R_{AA}^{\pi^0}$
at $2.76$ ATeV  cm energy. We update our earlier 2007 LHC predictions
in \cite{Abreu:2007kv,Horowitz:2007nq}, by extrapolating
the 2008 $1-\sigma$ PHENIX/RHIC constraints \cite{Adare:2008cg} 
of the opacity range at $\sqrt{s}=0.2$ ATeV
using the new 2.76 ATeV ALICE/LHC \cite{Aamodt:2010pb,Collaboration:2010cz} 
charged hadron rapidity density data, 
$dN_{ch}/d\eta=1601\pm 60$, in the $0-5\%$ most
central collisions and $35\pm 2$ in the $70-80\%$ peripheral collisions.

We note that in strong coupling AdS/CFT approaches to hard jet probes, 
the pQCD \highpt jet tomography theory 
is replaced by a gravity dual jet holographic model. 
That approach is based on the assumption that
the 't Hooft  coupling,
$\lambda\equiv 4\pi\alpha_s N_c$, as well as $N_c$ are 
large enough that an approximate 10D 
supergravity dual description of the dynamics may be used.  
Jet  quenching is then mapped into the problem of classical string drag 
 over a black brane horizon in an AdS curved spacetime with an isometry
group that is identical to the exact conformal symmetry group 
of the ${\cal N}=4$ Supersymmetric Yang-Mills (SYM) cousin of QCD. 
The thermally equilibrated strongly coupled supersymmetric QGP
is assumed dual to the black brane. 
In \cite{Noronha:2010zc}, it was shown
that with $\lambda\sim 20$ and $N_c=3$ ($\alpha_s\approx 0.5)$ 
AdS/CFT holography provides a remarkably robust analytic account of both
the single nonphotonic electron
(heavy quark jet) quenching as well as bulk
 elliptic flow data, a simultaneous description of which has remained out of reach of perturbative QCD methods.
However, the theoretical consistency of heavy quark jet holography
remains controversial in the literature (see, e.g.,
\cite{Hatta:2011gh}).
Light jet holography of $R_{AA}^{ch}$ 
is even more challenging.
So far, only schematic 
falling strings scenarios have been proposed to address
light quark/gluon  quenching observables \cite{Gubser:2008as,Chesler:2008uy}.  
However, high-$p_T$ elliptic moment $v_2(p_T)$ phenomenology
\cite{Wei:2009mj,Marquet:2009eq,Jia:2011pi,Betz:2011tu} 
appears to require nonlinear path length 
dependences, $L^{n\ge 3}$, more suggestive of falling string scenarios
than $L^{n\le 2}$ path dependences predicted by pQCD for static plasmas.
Future heavy flavor tagged jet observables \cite{Horowitz:2007su, Noronha:2010zc, Buzzatti:2010ck, Ficnar:2010rn} will help to discriminate between
competing jet holography vs.\ perturbative tomography models of jet attenuation.

\section{Jet Tomography: Qualitative}\label{section:qual}

One feature common to  both pQCD tomography and 
gravity dual holography is that both predict the energy loss per unit 
length, $dE/dL$, increases monotonically with the plasma density 
or temperature.  In this section we consider a generic analytic 
energy loss that can interpolate between a wide range of dynamical scenarios
and provide qualitative understanding of the quantitative numerical WHDG results presented in Section \ref{section:quant}.

Consider the following power law model for jet energy loss \cite{Betz:2011tu}
\begin{eqnarray}
\hspace*{-0.4cm}
\frac{dP}{d\tau} &=&
-{\kappa} P^a\tau^{b} T^{2-a+b}(x(\tau),\tau)
\label{Generic}
\end{eqnarray}
where $P(\tau)$ is the momentum (energy) of a massless jet 
passing through a  plasma with a local temperature field $T(x,\tau)$. 
The solution for an initial energy $P(0)$ 
and jet path, $x(\tau)$, a uniform static ``plasma brick'' of thickness $L$ is 
 \begin{eqnarray}
\hspace*{-0.4cm}
\frac{P(L)}{P(0)}
&=& 
\left( 1- \kappa (LT)^{1+b}\left( \frac{T}{P(0)}\right)^{1-a}\right)^{\frac{1}{1-a}}
\;\; .
\label{Genericeps}
\end{eqnarray}
We note that $\partial P(0)/\partial P(L)=(P(0)/P(L))^a$ is the Jacobian
of the transformation between $P(0)$ and $P(L)$.
The parameters $a$ and $b$ control the jet energy and path length
dependence energy (momentum) loss per unit length (time)
and fixes the power of the 
temperature or parton density $\rho\propto T^3$ dependence. 

The thermal stopping distance $L_T$, as defined by $P(L_T)=P_T=T$,
is then:
\begin{eqnarray}
\hspace*{-0.4cm}
L_T(P_0,T,a,b,c)&=&\left(\frac{P_0^{1-a}- T^{1-a}}{(1-a)\kappa T^{2-a+b}}\right)^{\frac{1}{1+b}}
\label{Stopping}
\end{eqnarray}
In pQCD, the opacity series WHDG/DGLV \cite{Gyulassy:2000fs,Gyulassy:2000er,Djordjevic:2003zk,Wicks:2005gt} for a massless parton  jet 
leads (in a static uniform plasma) to $dE_{GLV}/dL \approx - \kappa T^3 L^1 \log(E/T)$. Therefore $b=1$ (or $0$) in the LPM 
regime in static (or Bjorken expanding) plasmas.  
Because $E^{1/3}/\log(E)\approx 1\pm 0.1 $ in the range $5<E/T<200$, we can simulate pQCD light quark jet energy dependence with $(a\approx 1/3,b=1)$. 
The density dependence is then roughly
linear with 
$T^{8/3}\sim \rho^{8/9}$ in the static case.  

Another interesting limiting case 
in pQCD is for thick plasmas, where the deep LPM regime leads to the 
BDMS \cite{Baier:2001yt} formula, 
$dE_{BDMS}/dL=-\kappa E^{1/2}L^1 T^{5/2}$ that grows with density as $\rho^{5/6}$.
Increasing the density by a factor equal to the ratio of charged particle
multiplicities, $\approx 2.1$, from RHIC to LHC 
approximately doubles both the GLV and BDMS energy loss.

In distinction to pQCD, 
the stopping distance for light quark jets in the falling string holographic 
scenario was found in \cite{Chesler:2010bi} to be bounded by $L_T< \kappa E^{1/3}/T^{4/3}$.
This can be simulated via \eq{Stopping} by choosing 
$a>(2-b)/3$ in the range $a\in[0,1/3]$. 
The special case $b=2$ and $a=1/3$ 
is the one favored phenomenologically \cite{Adare:2010sp,Marquet:2009eq,Jia:2011pi,Betz:2011tu} 
by the high $p_T$ azimuthal anisotropy $v_2(p_T)$ data.
For that case, $dE_{AdS}/dL = - \kappa E^{1/3}L^2 T^{11/3}$. Thus 
the energy loss 
grows even faster with density, $\propto \rho^{11/9}$.  
We see that quite generally 
$dE/dL $ should increases significantly with density. 

For a more realistic Bjorken longitudinally expanding plasma of transverse 
thickness $L$ the density decreases approximately as
 $T^3(\tau)\propto (dN_{ch}/dy)/(\tau L^2)$. In this Bjorken brick case
 \begin{eqnarray}
\hspace*{-0.4cm}
\frac{P(L)}{P(0)}
&=& \left( 
1- \kappa' \frac{ (dN/dy)^{(2-a+b)/3}}{(L P(0))^{1-a}}\right)^{\frac{1}{1-a}}
\;\; .
\label{GenericBj}
\end{eqnarray}

To estimate the variation of $R_{AA}$ in this case 
consider the distribution of initial jet transverse momenta (at mid rapidity)
approximated as 
\begin{equation}
d\sigma(p_T)/dp_T \propto 1/p_T^{n(p_T,s)}
\end{equation}
 with $n(p_T,s)\equiv -d\log\big(d\sigma(s)\big)/d\log(p_T)$ given by the cm energy dependent 
parton spectral index computed from pQCD.

The nuclear modification factor for jet partons (neglecting hadron fragmentation) is simply a change of variables which leads 
with \eq{GenericBj} 
for a fixed centrality class to
 \begin{eqnarray}
\hspace*{-0.4cm}
R_{AA}(p_f;s,A)&=& \frac{\partial p_0}{\partial p_f} \frac{d\sigma(p_0(p_f))/dp}{d\sigma(p_f)/dp}\nonumber \\
&\approx & \left( 
1+ \kappa' \frac{ (dN/dy)^{(2-a+b)/3}}{(L p_f)^{1-a}}\right)^{\frac{a-n(p_f)}{1-a}}
\;\; .
\label{GenericBjRAA}
\end{eqnarray}
Given a model specified by $(a,b)$ 
the coupling parameter $\kappa'(a,b)$ can be fixed by fitting
RHIC $R_{AA}$ at one reference momentum, e.g.\ $p_T=10$ GeV/c.

\begin{figure}
\includegraphics[width=\columnwidth]{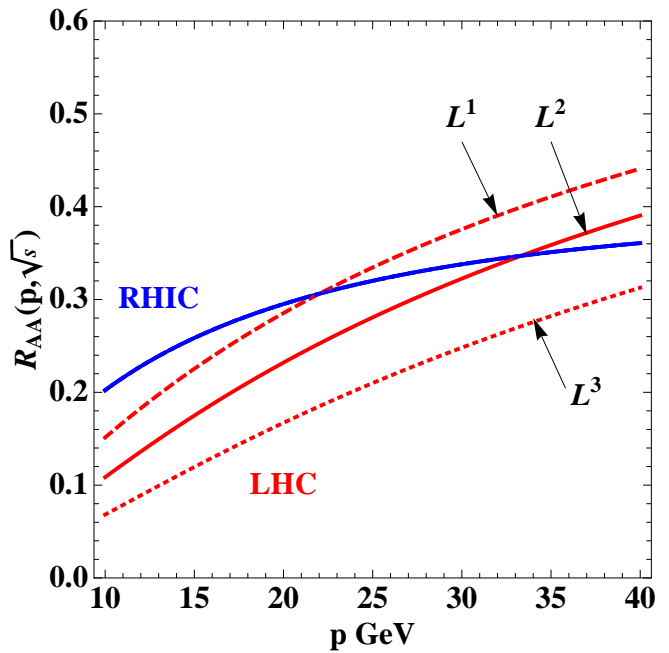} \\
\caption{\label{fig:toy}Schematic nuclear modification factor
model \eq{GenericBjRAA} 
for a ``Bjorken plasma brick''
 normalized to $R_{AA}(p_T=10)=0.2$ for central RHIC (blue) assuming $a=1/3$. 
The momentum dependence at RHIC is independent of $b$ once the reference
normalization at $p_T=10$ GeV/c is fixed. The red curves for LHC are evaluated 
with initial rapidity density scaled up  by a factor 2.0 relative to RHIC.
The $b=0,\,1,\,2$ dependent  extrapolations to LHC (red dashed, solid, dotted)
are labeled by the equivalent $L^b$ path length 
dependence of the total energy loss in the static plasma limit. 
}
\end{figure}

We show in \fig{fig:toy} the evolution of this simple analytic model of
$R_{AA}$ from RHIC to LHC
for the case $a=1/3$ energy loss and for different $b=0,1,2$ that 
correspond to path length scaling for  elastic, inelastic and
AdS falling strings scenarios. For this qualitative plot, we 
take  spectral indices (see \fig{fig:gluefraction} of the next section) as 
$n_{RHIC}\equiv n_1(p_T)\approx 5.5+(p_T-10)/10$ that rises by 1 unit 
in the $p_T=10-20$ range at RHIC 0.2 ATeV, while $n_{LHC}\equiv n_2\approx 4.5$ approximately independent of $p_T$ at 
LHC 2.76 ATeV. We fix $\kappa'$ for each $b$ 
by demanding $R_{AA}(p_T=10,\,dN/dy=1000)=0.2$.
The RHIC blue curve is therefore independent of the $b$ parameter.
For larger $a$ the blue curve flattens, while for smaller $a$ the curve
rises with $p_T$ more rapidly. The $a=1/3$ rise is within the large error
band of the PHENIX data shown in \fig{fig:RAAnow}. 

It is important to note that if there were no density dependence
of energy loss, then just because the spectral index
decreases from $\sim 5.5$ to 4.5, the $R_{AA}(LHC)$ would be less quenched 
by $\sim 70\%$ at $p_T=10$ GeV/c at LHC! However,
the approximate doubling of the initial density (and hence $dE/dL$) at 
LHC relative to RHIC 
results in halving $R_{AA}$ at $p_T=10$ GeV/c relative to RHIC.
In addition, the reduction of the fractional energy loss with increasing
$p_T$ with $a=1/3$ causes $R_{AA}$ to rise with $p_T$ at LHC while
at RHIC the $p_T$ dependent increase  of the spectral index
compensates this natural rise. 
 We will confirm these generic features in the next section with detailed
WHDG numerical calculations.

With this simple analytic model we can also easily explore the sensitivity
of the LHC extrapolation to the path length dependence parameter $b$.
The $b$ parameter influences mainly the scaling exponent of $dN/dy$ in the Bjorken expanding case.
We see that decreasing $b$ to 0  (simulating perturbative elastic energy loss)
decreases the difference between RHIC and LHC at $p_T=10$ GeV/c but the change in $p_T$ slope is similar. 
For $b=2$ (simulating an $L^3$ path dependence suggested by
AdS falling strings) 
predicts a significantly larger density dependence from RHIC to LHC.
The cross over momentum  where $R_{AA}$ at RHIC and LHC are equal
increases monotonically with $b$. 
The absolute value of $R_{AA}$ and its $p_T$ slope therefore 
provide strong constraints on the $a,b$ parameters 
and therefore  on the initial  density or  $dN/dy$ dependence, 
$\rho^{(2-a+b)/3}$  None of the parametric models 
predicts similar nuclear modification factors
at RHIC and LHC above $p_T>10$ GeV/c.

\section{Jet Tomography: Quantitative}\label{section:quant}

The extraction of tomographic (density dependent)
information from \raacomma, and suppression observables in general, is
complicated by many competing physics processes---some of which are
interesting in their own right.
Of greatest import for our discussion is the initial phase space distribution of
\highpt quarks and gluons in both coordinate and momentum space.  At
RHIC, extensive reference spectra from p + p collisions at the same $\sqrt{s}$
as in Au + Au collisions allow for a precise calibration
of \raa and
constrain experimentally the initial quark and gluon spectra.
The control d + Au data were essential to calibrate
the magnitude of the initial state nuclear distortions of the initial jet spectra
due to gluon saturation/shadowing physics \cite{Gyulassy:2004zy,McLerran:2008pe,Eskola:2009uj,Eskola:2010jh,Paukkunen:2010qi}.
At midrapidity the initial \highpt
partonic spectrum in the $(x,Q^2)\sim (0.1,100)-(0.2,400)$ kinematic
range was not found to deviate significantly from the reference
 p + p pdfs (see Fig.\ 1 of \cite{Paukkunen:2010qi}). Direct photon measurements further confirmed
that the binary collision scaling 
based on  Glauber diffuse nuclear reaction geometry  
can be used to understand the nuclear number and impact parameter
dependence of hard processes \cite{Adler:2005ig}.

Comparable control measurements do not yet exist at LHC, thereby severely
limiting the strength of conclusions that can be 
drawn from the first  \raa data \cite{Aamodt:2010jd}. 
The absence of reference p + p data leads alone to about a factor
$\sim 2$,  $p_T$-dependent systematic uncertainty in the normalization of
the reported \raacomma.  This uncertainty is shown for the 0-5\%
centrality data by the yellow band in \fig{fig:RAAnow}.  Second, the absence
of control p + Pb as yet makes it impossible to deconvolute
\emph{initial state} nuclear suppression 
of \highpt particles due to
small x gluon saturation, or Color Glass Condensate (CGC)  \cite{Blaizot:1987nc,McLerran:1993ni,Iancu:2003xm,Gyulassy:2004zy,Weigert:2005us,JalilianMarian:2005jf,McLerran:2008pe,Albacete:2010bs}, effects at LHC in the $Q^2>100$ GeV$^2$ range.
Because the fractional momenta relevant for midrapidity jet production
at LHC are 10 times smaller at a given $p_T$ than at RHIC
strong initial state suppression of the nuclear/gluon structure
has been predicted in \cite{Albacete:2010bs} at LHC.
 
In this work, we assume the absence of significant initial-state suppression in the $p_T\sim 10-20$ GeV/c kinematic region corresponding to
$(x,\,Q^2)\sim (0.01-0.02,\,100-400$ GeV$^2)$. This is consistent with the DGLAP
$Q^2$ evolution of global fits to nuclear pdfs 
(see Fig.1 of \cite{Paukkunen:2010qi}). 
The first ATLAS measurement \cite{Collaboration:2010px} of Z boson 
candidates is also consistent with unshadowed
binary scaling of jets in the $x\sim 0.05$ kinematic range.
Future direct photon
measurements at LHC will provide additional 
control over initial state
shadowing/CGC effects. 

\begin{figure}
\includegraphics[width=\columnwidth]{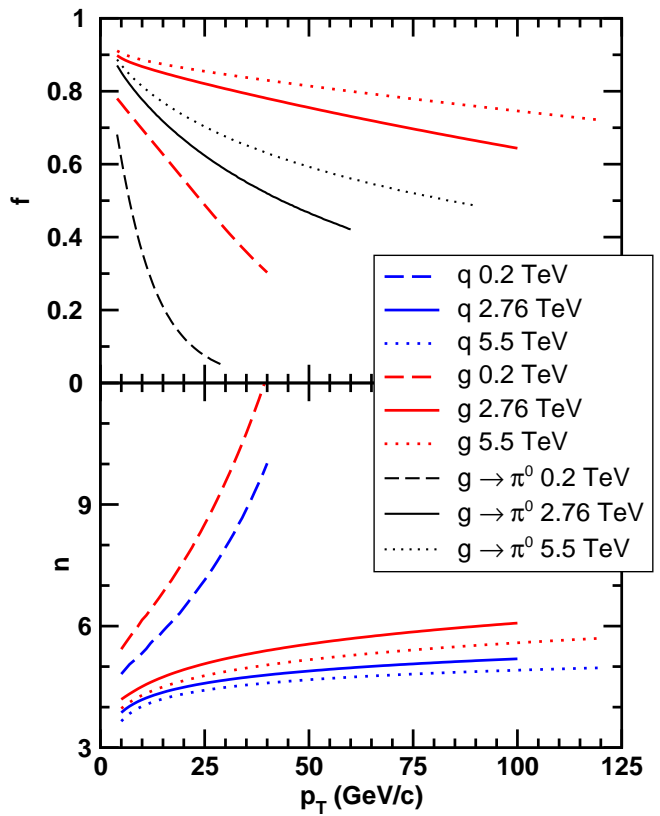} \\
\caption{\label{fig:gluefraction}(Top) Fraction of jets that are gluons
as a function of jet \pt from LO pQCD \cite{XNcode} for $\sqrt{s}=0.2,\, 
2.76,$ and 5.5 TeV shown in red. Black curves show fraction of pions fragmenting
from gluons at final pion $p_T$.   
(Bottom) Partonic spectral index, $n(p_T)= -d\log(dN/dydp_T)/d\log(p_T)$, 
of initial $y=0$ gluon (red) and light quark (blue) jets 
compared at RHIC and LHC in $p+p$.}
\end{figure}

The new ALICE data show features that appear
strikingly similar to expectations based on pQCD energy
loss: in particular \raa rises significantly at LHC as a function of \pt
rather than the observed flatness within errors at RHIC.  In \cite{Vitev:2002pf}
the predicted stronger rise as a function of momentum at LHC 
can be understood from the qualitative model in Section \ref{section:qual} and
the also from the following even simpler schematic model: 
The fractional energy loss of a
\highpt parton decreases in pQCD with
 momentum as $\epsilon\sim\log(\eqnpt)/\eqnpt$.  The 
final momentum, $\eqnpt^f$, is related to the initial momentum,
$\eqnpt^i$, by $\eqnpt^f=(1-\epsilon)\eqnpt^i$.  For particle
production distributions approximated by a power law,
$dN/d\eqnpt\sim\eqnpt^{-n}$, the nuclear modification
factor  $\eqnraa \sim \langle
(1-\epsilon)^{n-1} \rangle$.  The suppression at RHIC is flater than at 
LHC due to an accidental 
cancellation between 1) the fraction of \highpt gluons to quarks (see
the upper panel of \fig{fig:gluefraction}), 2) the hardening of the
production spectrum as a function of \pt (see the lower panel of
\fig{fig:gluefraction}), and 3) the decrease in energy loss as a
function of \pt (see \fig{fig:DeltaE}).
 
As shown in the lower panel of \fig{fig:gluefraction} at LHC, the
production spectrum is much harder (smaller $\sim$ 
constant spectral index) 
than at RHIC. For a constant $n(p_T)$ the decrease of fractional energy loss
with \pt would lead to an \raa that
increases with \ptcomma.  One can see from \fig{fig:RAAnow} that 
the WHDG prediction rises with \pt similar to the ALICE data.

As one can see from the top
panel of \fig{fig:gluefraction}, also in contrast to RHIC, LO pQCD predicts
that hard jets at LHC are predominately gluons to much higher $p_T$.
 Na\"ively then, with larger density, larger
medium size (for Pb vs.\ Au), and the greater preponderance of
gluons with $9/4$ enhanced energy loss relative to quark jets, 
should lead to an \raa suppressed much below that
seen at RHIC.  However, the smaller and flatter spectral indices of both quarks and gluons compensates in the other direction.

The first numerical GLV prediction for $R_{AA}^{\pi^0}$ in 2002 for 5.5 ATeV Pb + Pb collisions LHC 
including only radiative energy loss
was given in Fig.\ 3 of \cite{Vitev:2002pf} and overlaps well with the yellow
ALICE systematic error band in \fig{fig:RAAnow}.  The first predicted WHDG \raa for LHC (see Fig.\ 83 of \cite{Abreu:2007kv}) 
including  elastic energy loss as well as
 radiative energy loss 
and also realistic geometric jet path fluctuations 
accidentally remained close to Vitev's original GLV prediction.
Our currently updated WHDG study incorporates
the observed 2.2 times increase in charged 
particle rapidity density by ALICE in Pb+Pb 
2.76 ATeV collisions to extrapolate  
the most recent $1-\sigma$ uncertainty band of initial sQGP 
densities compared to high statistics PHENIX data from RHIC: this is the blue band in \fig{fig:RAAnow}, which shows a significant underprediction of the ALICE data.

We discuss in more detail below a range of theoretical and experimental 
issues that need further attention as jet tomography advances toward
a more quantitative level.
 The RHIC constrained  WHDG results for \raa
appear consistent with the ALICE data and the earlier estimates
within the present very large error band
 due to the unknown p + p baseline at this energy. 
While stronger conclusions
are not warranted at this time, 
perturbative QCD tomography is at least
consistent with the 
striking positive slope $p_T$ dependence of the 
nuclear modification light parton jet fragments at LHC. The change of the $p_T$
slope, $dR_{AA}/dp_T$, 
from negative at SPS to approximately zero at RHIC and finally to positive 
at LHC is a critical consistency test for the pQCD framework 
that jet holographic approaches must (eventually) also be able to pass.

\section{From Energy Loss to \texorpdfstring{$\eqnraa$}{RAA}}
\subsection{0-5\% Central}
\begin{figure}
\includegraphics[width=\columnwidth]{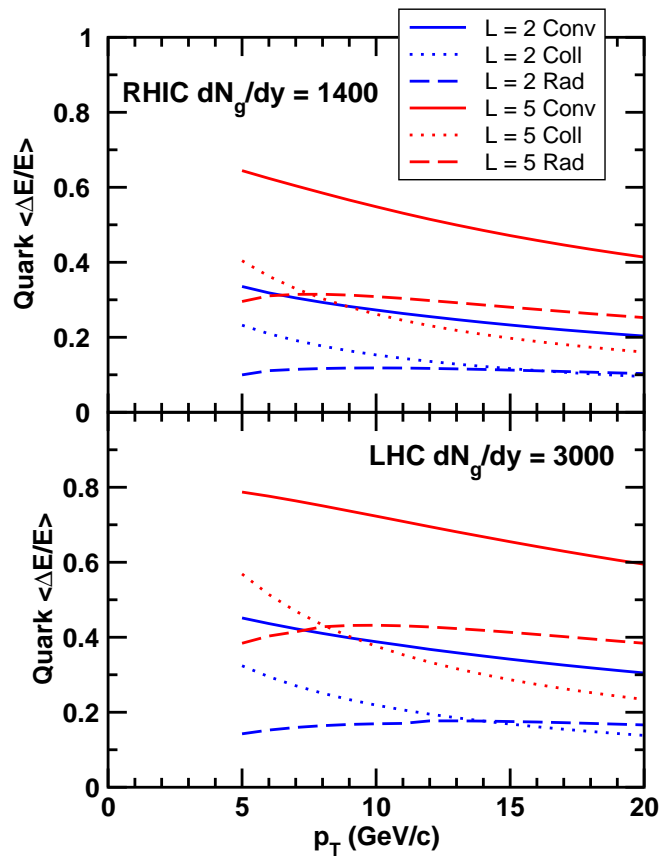} \\
\caption{\label{fig:DeltaE}Average fractional energy loss
  $\langle\epsilon\rangle=\langle\Delta E/E\rangle$ for a light quark
  of initial momentum \pt in a Bjorken-expanding brick of plasma of
  length 2 fm (blue) and 5 fm (red) for RHIC (top) and LHC (bottom)
  conditions from the WHDG energy loss model \cite{Wicks:2005gt}.
  $\langle\epsilon\rangle$ is shown for purely radiative energy loss
  (Rad), purely collisional energy loss (Coll), and the incoherent
  convolution of the two (Conv).  Elastic energy loss is a crucial
  contribution to total energy loss at both RHIC and LHC.}
\end{figure}

The WHDG energy loss calculation
described in \cite{Wicks:2005gt,TECHQM} uses the first order in opacity
radiative energy loss 
and Braaten-Thoma pQCD collisional partonic energy loss. 
The model assumes  Debye-screened color scattering centers, and 
one loop screening mass $\mu= g T$. Both light quarks and gluons have
one loop medium-induced thermal masses of order $\mu$.  As noted previously, a
generic feature of this pQCD radiative energy loss is that the
fraction of final momentum to initial momentum of the parent parton,
$\epsilon = p_T^f/p_T^i$, scales as $\epsilon\sim \log(p_T)/p_T$ as can be seen 
\fig{fig:DeltaE}.  Of crucial importance in this formalism is the quantum
interference between the hard vacuum radiation due to the initial
creation of a hard parton and the radiation induced by scatterings in
the medium. 

The redistribution of energy via collisional processes is found
through the use of the fluctuation-dissipation theorem with the mean
loss given by the thermal field theory calculation of Braaten and
Thoma\cite{Braaten:1991we}.  
This redistribution of energy can lead to the \highpt
parton either losing or gaining energy, which is taken into
account.  The simple Gaussian approximation due to the
fluctuation-dissipation theorem is a simplification
that future more detailed but unwieldy
elastic energy loss calculations can improve
\cite{Wicks:2008zz}. One can see in
\fig{fig:DeltaE} that, with these assumptions and at medium densities
appropriate for RHIC and LHC, at moderate $\eqnpt\lesssim20$ GeV
elastic energy loss is as important as inelastic at both RHIC and LHC.

We note in passing that the production points of the high momentum
particles are distributed according to the binary distribution
$T_{AA}$; the medium density is assumed proportional to the
participant density given by the Glauber model, and 1-D
Bjorken-expansion is approximately included by evaluating the density
at a time one half the path length.

\raa is defined as the ratio of observed spectra in A + A collisions
for a given centrality divided by the observed spectrum in p + p
collisions scaled by the number of binary collisions at the given
centrality.  ALICE measured the \raa of hadrons using the
non-invariant spectra, \be
\eqnraa^{h^++h^-}(\eqnpt,\,b)=\frac{dN^{h^++h^-}_{AA}/d\eqnpt
  dy}{N_{\rm coll}(b)dN^{h^++h^-}_{pp}/d\eqnpt dy}, \ee
When one assumes
that the production points of hard particles is proportional to the
\taa distribution, where $N_{\rm coll}=\int d^2\wv{x} \, T_{\rm
  AA}(\wv{x})$, \ncoll drops out of the theoretical calculation.
However it is important to note that the values of \ncoll that we find
using the optical Glauber model with the same parameters listed in
\cite{Aamodt:2010jd} significantly disagree with those used by ALICE, found using a
Monte-Carlo approach; we quantify this discrepancy in \tab{tab:ncoll}.
We will come back to this difference when we discuss $R_{\rm cp}$
below.

\begin{table}[!htbp]
\begin{tabular}{|c|c|c|c|c|}
\hline
& ALICE & ALICE & Opt.\ Gl.\ & \\
Cent.\ & \ncoll  & Rel. Unc. & \ncoll & \% Diff \\\hline
0-5\% & $1690\pm131$ & $\sim8\%$ & 1710 & $\sim1\%$ \\
70-80\% & $15.7\pm0.7$ & $\sim5\%$ & 12.6 & $\sim25\%$ \\\hline
\end{tabular}
\caption{\label{tab:ncoll}Values of \ncoll used by ALICE and those
  found using an optical Glauber model with a Woods-Saxon geometry and
  inelastic cross section identical to those used by ALICE in their
  Monte-Carlo calculation. Cf.\ to the uncertainties shown in
  \fig{fig:RCP}.}
\end{table}

\begin{figure}
\includegraphics[width=\columnwidth]{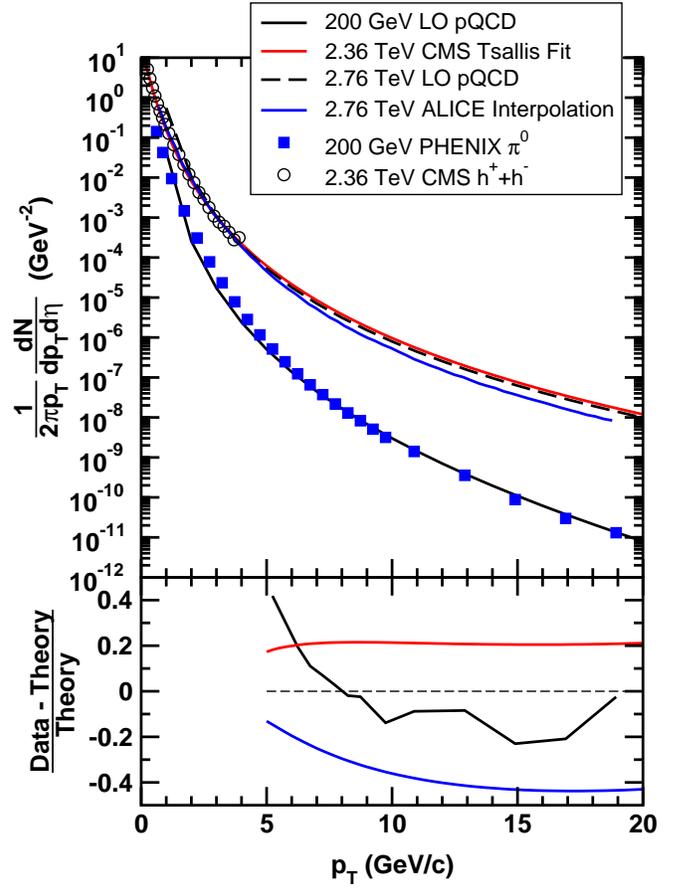}
\caption{\label{fig:spectra}(Top) Hadron production spectrum data in p
  + p collisions from PHENIX at $\sqrt{s} = 0.2$ TeV \cite{Adare:2007dg} and from
  CMS at $\sqrt{s}=2.36$ TeV \cite{Khachatryan:2010xs}; the Tsallis distribution that is
  a best fit to the low $p_T$ CMS data is shown in red.  Also in the figure is
  the LO pQCD calculation at $\sqrt{s} = 0.2$ TeV (black) and
  $\sqrt{s} = 2.76$ TeV (dashed black).  The interpolation 
spectrum used by ALICE to
  normalize \raa at $\sqrt{s} = 2.76$ TeV is shown in blue.  (Bottom)
  The PHENIX data normalized to $\sqrt{s}=0.2$ TeV LO pQCD (black),
  the CMS Tsallis fit at $\sqrt{s}=2.36$ TeV normalized to LO pQCD at
  $\sqrt{s}=2.76$ TeV (red), and the ALICE interpolation normalized to
  LO pQCD at $\sqrt{s}=2.76$ TeV (blue).  Despite a larger $\sqrt{s}$,
  note that the ALICE interpolation is a factor $\gtrsim2$ smaller than the
  CMS Tsallis fit, emphasizing the obvious need for 2.76 pp reference data.}
\end{figure}

In order to make contact with the experimentally observed hadrons, the
partonic energy loss described above must be convolved with a partonic
production spectrum and a set of fragmentation functions (FFs).  We
compute partonic suppressions via $R_{\rm AA}^{q,g} = \langle
(1-\epsilon(\wv{x},\,\phi))^{n^{q,g}(\eqnpt)-1} \rangle$, where we
take $n^{q,g}=d\log(dN^{q,g}/d\eqnpt)/d\log(\eqnpt)$ to 
approximate the power law for the partonic production spectrum,
and $\langle\cdots\rangle$ represents an averaging over the geometry
for a given centrality class.  Then
\begin{multline}
\label{eq:RAA}
R_{\rm AA}^h(\eqnpt;\,b) \\ = \int \frac{dz}{z}
\frac{dN^q}{d\eqnpt}\bigl(\frac{\eqnpt}{z}\bigr)R_{\rm
  AA}^q\bigl(\frac{\eqnpt}{z};\,b\bigr)D^{q\rightarrow h}\bigl(z\bigr)
\, + \, q\rightarrow g\Big/ \\ \int \frac{dz}{z}
\frac{dN^q}{d\eqnpt}\bigl(\frac{\eqnpt}{z}\bigr)D^{q\rightarrow
  h}\bigl(z\bigr) \, + \, q\rightarrow g
\end{multline}
We use a LO pQCD code \cite{XNcode} with a $K$ factor of 3,
CTEQ5L parton distribution functions \cite{Lai:1999wy} evaluated at
$Q=\eqnpt/2$, and KKP fragmentation functions \cite{Kniehl:2000fe}. 
It is important to emphasize that the factor of two systematic uncertainty in
the $K$ factor drops out of the ratio above.  
The produced neutral pion spectra expected for p + p collisions at
RHIC and LHC energies using this LO pQCD procedure are shown in the
top panel of \fig{fig:spectra}.  The LO pQCD spectrum at
$\sqrt{s}=0.2$ TeV agrees well with the PHENIX data to within
$\pm20\%$.  On the other hand we find that our LO pQCD spectrum
with $K=3$ evaluated at $\sqrt{s}=2.76$ TeV systematically overpredicts the
spectrum used by ALICE by a nearly \ptcomma-independent factor of
$\sim2$.  We will come back to the  absolute cross section 
normalization differences below.

One of the key ingredients in an energy loss calculation is the
density of the medium through which a fast parton propagates.  We
assume the transverse coordinate dependence of the density of the
quark-gluon plasma medium is proportional to the participant density,
$dN_g/d^2\wv{x}dy\propto\rho_{\rm part}$.  The proportionality
constant that connects these two quantities is precisely the lowest
order tomographic information we can deduce by comparing energy loss
calculations with data.  The PHENIX collaboration extracted this
constant, reported as $dN_g/dy$, via a rigorous statistical analysis
comparing the WHDG energy loss to the 0-5\% central RHIC $R_{\rm
  AA}^{\pi^0}(\eqnpt)$ data \cite{Adare:2008cg}; it was found that the
best fit value and one standard deviation uncertainty are $dN_g/dy =
1400^{+200}_{-375}$ for a fixed $\alpha_s=0.3$.  Once this constant is fixed and we make the
assumption that the quark-gluon plasma medium density scales with the
observed number of charged hadrons, the QGP medium density at LHC is
predicted.  ALICE reported that the 0-5\% central value of $(dN_{\rm
  ch}/d\eta)\big/(N_{\rm part}/2)$ increased by a factor of 2.2 from
RHIC \cite{Aamodt:2010pb};
the scaled density in 0-5\% central collisions at $\sqrt{s_{NN}}=2.76$ TeV is then $dN_g/dy = 3000^{+500}_{-800}$.  

With the QGP medium density at LHC so fixed, 
the suppression from the WHDG
model is a \emph{constrained prediction}; i.e.\ there are no free
parameters in our calculation.  We compare the resulting suppression
for 0-5\% central $R_{\rm AA}^{\pi^0}(\eqnpt)$ in \fig{fig:RAAnow} to
the 0-5\% central ALICE charged hadron suppression.  The results from
the one standard deviation uncertainty in $dN_g/dy$ are represented by
thin blue lines; a light blue shaded region connects the two,
denoting the uncertainty in the theoretical calculation due to the
uncertainty of the extracted medium parameter from RHIC data.  The
dashed blue line represents $R_{\rm AA}^{\pi^0}(\eqnpt)$ for the 
best 
fit value of $dN_g/dy = 3000$.  One can see from the figure that the
perturbative calculation qualitatively describes the increase in \raa
as a function of \ptcomma, as we expected.  Also our pocket formula
from above describes the \lowpt normalization of the WHDG \raa results
rather well.  The small, $\lesssim0.1$ value of \raa at \lowpt
demonstrates again that the WHDG energy loss model is not fragile.
The realistic distribution of production points and medium density
means there is no (appreciable) corona of jets that emerge unmodified;
there is no lower bound to the theoretical value of \raacomma.

On the other hand, the very low normalization of the WHDG \raa
seriously underpredicts the central values of the ALICE data.  As
there is no measured p + p baseline at $\sqrt{s}=2.76$ TeV at LHC yet,
ALICE reports a very large uncertainty in \raa due to the uncertainty
in the interpolated p + p baseline used.  The result is that the 0-5\%
central WHDG energy loss calculation and the ALICE data agree at the
edge of their respective reported 1-$\sigma$ uncertainties.  It is
worth emphasizing again, that 
while the WHDG \raa prediction is independent of the normalization of the
p+p reference, the ALICE \raa is sensitive to the unmeasured p+p
reference that has an intrinsic theoretical factor of $\sim$ two  
systematic uncertainty
at both LO and NLO level.

In addition to the theoretical uncertainty due to the finite precision
extraction of a medium density at RHIC, there are also systematic
uncertainties in the theoretical calculation due to simplifying
approximations made in the derivation of energy loss formulae.  Some
of these uncertainties have been quantified \cite{Horowitz:2009eb}, and it turns
out that the uncertainty due to the collinear approximation is in fact
quite large: an exploration of this systematic theoretical uncertainty
yields a factor of 3 uncertainty in the extracted medium density at
RHIC when energy loss is assumed to be radiative only.  Given this
large uncertainty, one might na\"ively expect a similarly large
uncertainty in the WHDG predictions for \raa at LHC.  However this
expectation is incorrect for two reasons: first, the uncertainty due
to the collinear approximation decreases significantly when elastic
energy loss is included \cite{Horowitz:2009eb}. Second, 
observables such as $v_2(p_T)$---once the reference opacity is fixed---are found numerically not to depend
much on the specifics of the explored collinear approximation
uncertainty \cite{Horowitz:2011}.
The theoretical systematic
uncertainties of the elastic energy loss contributions are
less well known and call for more scrutiny. In \cite{Wicks:2008zz}
it was shown that, contrary to often assumed multi soft Gaussian transverse
elastic diffusion approximation, the large momentum transfer power law tails 
with $q\sim 10 T\sim 2$ GeV momentum exchange (as included in WHDG through 
the $\log(E/T)$ enhancement of elastic energy loss),
cannot be ignored as assumed in many other models.  
In order to quantify the size of this collinear  radiation
approximation systematic uncertainty we use the inelastic energy loss formula derived in Minkowski coordinates and take the maximum angle of emission for radiation to be $\theta_{\rm max}=\pi/2$ (for a more detailed discussion, see \cite{Horowitz:2009eb}).  
The best fit value for the average medium density constrained by central RHIC data is $dN_g/dy=1400$ \cite{Horowitz:2009eb}, which extrapolates to $dN_g/dy=2800$.  The purple curves in \fig{fig:RAAnow} and denoted by ``$x=x_E$'' in \fig{fig:RCP} shows the values of $R_{\rm AA}(\eqnpt)$ that result.  That the curve lies within the WHDG uncertainty blue band due to the 1-$\sigma$ medium parameter extraction at RHIC confirms that the systematic theoretical uncertainty due to the collinear approximation is small for \raa at LHC once the medium parameter is fixed to RHIC data; in particular, the WHDG predictions and the ALICE data are in quantitative tension within the combined experimental and currently quantitatively explored theoretical uncertainties.

\subsection{70-80\% Peripheral and \texorpdfstring{$\eqnrcp$}{Rcp}}

Continuing with the assumption that the QGP medium density scales with
the number of charged hadrons we may make another \emph{constrained
  prediction} from WHDG for the \raa of the 70-80\% centrality class. From the ALICE centrality-dependent multiplicity data
\cite{Collaboration:2010cz} we find that $dN_g/dy(70-80\%) =
66^{+10}_{-17}$, and $dN_g/dy = 62$ for the $x=x_E$ calculation.
These are admittedly very small densities distributed over a small
region, and it is not clear that a QGP medium forms in these highly
peripheral heavy ion collisions.  Nevertheless we have assumed that
our energy loss formalism is valid in both deconfined and confined
media; it is not unreasonable to compare our results to the data, and
our calculation and the ALICE results shown in \fig{fig:RAAnow}.

\begin{figure}
\includegraphics[width=\columnwidth]{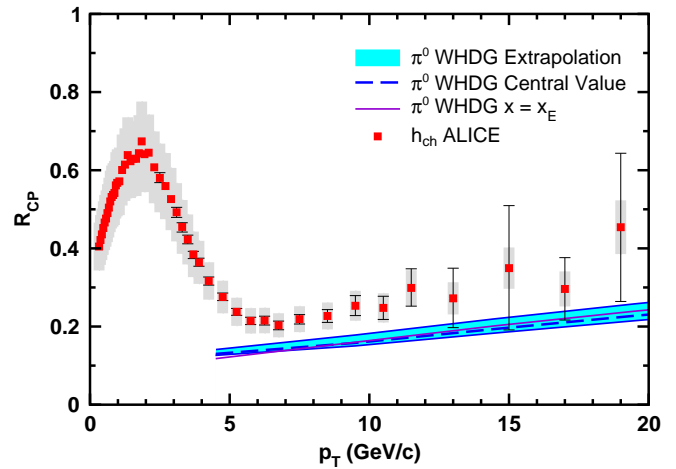}
\caption{\label{fig:RCP}$R_{\rm cp}$, the ratio of 0-5\% central
  \raapt to 70-80\% peripheral \raaptcomma, from ALICE and our WHDG
  calculations.  Experimental errors were added in quadrature, with
  those uncertainties due to the unknown p + p baseline cancelled.}
\end{figure}

Since there is so much uncertainty in the p + p baseline spectrum, we
find it useful to examine \rcpcomma, which is the ratio of central
\raa to peripheral \raacomma, in this case 
\be
R_{cp}(\eqnpt) =\frac{ R_{AA}(\eqnpt; \, 0-5\%)}{R_{AA}(\eqnpt; \, 70-80\%)}.
\ee
We plot both the experimental
values and our calculation in \fig{fig:RCP}.  Care was taken to
propagate the relative systematic and statistical errors in
quadrature.  Some component of the systematic uncertainty shown in the
original ALICE figures \cite{Aamodt:2010jd} is due to a \ptcomma-dependent
uncertainty from the unknown p + p reference spectrum, which cancels
in the $R_{cp}$ ratio.  The ALICE paper \cite{Aamodt:2010jd} quotes the
systematic uncertainties not due to the p + p spectrum as 5-7\% and
8-10\% in the central and peripheral bins, respectively.  As a
conservative estimate we take the upper values and add (in quadrature)
the relative uncertainty due to the \ncoll normalization; this
procedure yields a systematic uncertainty of 15\%, represented by the
gray bands in \fig{fig:RCP}.  Again the light blue band represents the
theoretical uncertainty due to the extraction of the medium density at
RHIC, with the dashed blue curve representing the best fit value, and
the thin purple curve represents the result when using the $x=x_E$
calculation discussed above.  

The theoretical calculation reproduces
the observed \ptcomma-dependence of \rcp quite well.  The theory
results are systematically about 2 standard deviations of systematic
uncertainty below the data; however one should note that much of this
systematic error is correlated and \ptcomma-independent, and a smaller
peripheral \ncollcomma, as suggested by the results from an optical
Glauber calculation, would yield an experimental \rcp suppression
significantly closer to our prediction.  Precise observations of
direct photons and Z bosons should help reduce this possible extra
uncertainty on the number of binary collisions in highly peripheral
events.  Additionally, that the optical Glauber results deviate so significantly from the Monte Carlo results in the 70-80\% centrality bin provides a quantitative feel for the importance of geometry fluctuations for these highly peripheral collisions; these possibly large geometry fluctuations are not taken into account in the energy loss calculations presented here.  Nevertheless, we find that this large discrepancy is a
challenge to the pQCD paradigm assumed by the WHDG energy loss
calculation.

\begin{figure}
\includegraphics[width=\columnwidth]{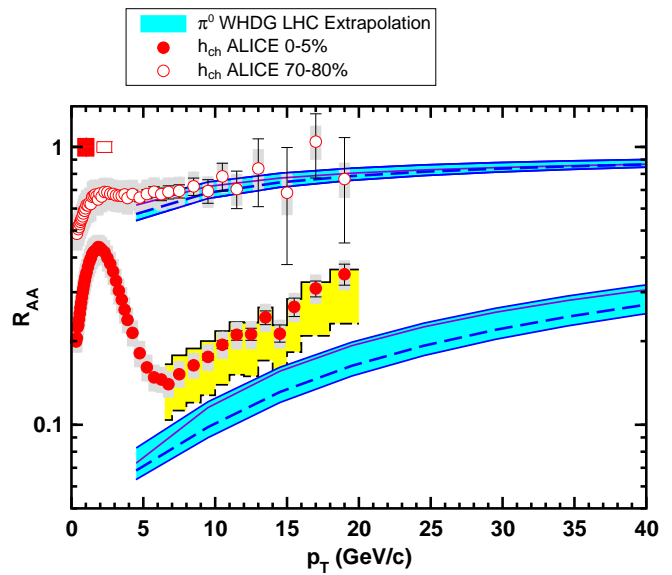}
\caption{\label{fig:RAA276high}Predictions of $R_{AA}^{\pi^0}$ for Pb + Pb collisions at 2.76 ATeV from the WHDG energy loss model out to $p_T=40$ GeV/c compared to current ALICE measurements of $R_{AA}^{h_{ch}}$ \cite{Aamodt:2010jd}.}
\end{figure}

\begin{figure}[!tp]
\includegraphics[width=\columnwidth]{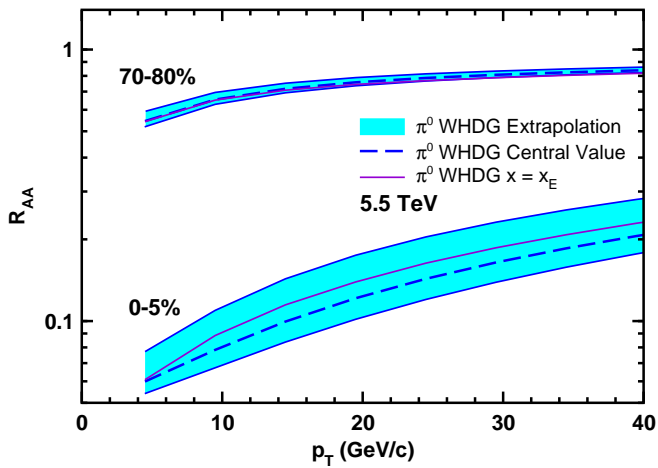}
\caption{\label{fig:RAAfuture}Central and peripheral $R_{\rm
    AA}^{\pi^0}$ at $\sqrt{s_{NN}}=5.5$ TeV from the WHDG energy loss
  model.}
\end{figure}

In \fig{fig:RAA276high} we present our predictions for $R_{AA}^{\pi^0}$ measured in Pb + Pb collisions at 2.76 at LHC out to $p_T=40$ GeV/c.  \fig{fig:RAAfuture} shows our predictions for central and peripheral suppression of neutral pions at $\sqrt{s_{NN}}=5.5$ TeV.  
Assuming that charged particle multiplicity scales as
$(s_{NN})^{0.15}$, as ALICE reported by fitting current world data
\cite{Aamodt:2010pb}, and assuming medium density scales with the
observed charged particle multiplicity, we have that at
$\sqrt{s_{NN}}=5.5$ TeV the extrapolation from the RHIC extraction is
$dN_g/dy = 3700^{+500}_{-1000}$; the light blue band in the figure
represents the predicted suppression based on these medium densities,
with the dashed blue curve representing the most likely value.  The
thin purple curve represents the result when using the $x=x_E$
prescription and an extrapolation of $dN_g/dy = 3500$.  
We note that these constrained predictions for 5.5 ATeV are qualitatively similar
but differ in detail from our earlier
2007 predictions that
assumed a smaller medium density range, $dN_g/dy =
1700-2900$ \cite{Horowitz:2007nq}. These updated 
WHDG constrained predictions are qualitatively 
similar also to the predictions 
 from 2002 GLV \cite{Vitev:2002pf} that assumed smaller opacities at 5.5 ATeV
 but neglected the competing effects due to elastic and radiative energy loss
as well as path length fluctuations included in WHDG.  Similar pertubative overquenching was also noted in other energy loss model approximations in \cite{Che:2011vt,Renk:2011gj,Levai:2011qm}. As emphasized in Section \ref{section:qual}
overquenching is a generic robust prediction of density dependent dE/dL models.

\section{Conclusions and Discussion}
In this paper we compared $\pi^0$ \raapt and \rcppt from the WHDG energy loss
model \cite{Wicks:2005gt} at $\sqrt{s_{NN}}=2.76$ TeV and the recent
ALICE data on charged hadron suppression at LHC \cite{Aamodt:2010jd}.  The WHDG
model includes both radiative and collisional channels and jet path length fluctuations. 
We found
that at the momentum range currently probed at LHC, collisional energy
loss is as important as radiative energy loss and cannot be neglected.  Previous work \cite{Horowitz:2010yi} showed that
even at top LHC energies and $\eqnpt\sim250$ GeV, elastic energy loss
makes up a sizable fraction, $\sim25\%$, of the total energy loss in the HTL
pQCD picture of the sQGP.
The results we present assume that the HTL QGP medium density scales with
the global charged particle multiplicity.  Our results are true 
predictions based on rigorous statistical constraints
of the medium density using  WHDG at RHIC to the QGP conditions
at 2.76 central Pb+Pb as constrained by global $dN_{ch}/dy$ from ALICE:
\emph{there are no free parameters}.  While unconstrained models without elastic energy and path length fluctuations can fit any \raa by adjusting the opacity, consistency with extensive 
RHIC jet quenching data leads in WHDG to a prediction of overquenching
relative to the first ALICE \raa data.

We find that our results show a qualitative agreement with the
momentum dependence of the central ALICE data and a quantitative
agreement with the normalization and \ptcomma-dependence of the
peripheral ALICE results, within the reported uncertainties.  Under
our assumption of the scaling of the medium density, though, we find a
quantitative disagreement with the normalization of the suppression for
the reported central values of the ALICE data for 0-5\% central Pb + Pb collisions.
Unfortunately, the very large systematic uncertainty in the ALICE data due
to the unmeasured base p + p reference precludes strong conclusions
at this time. Our calculated \raa is free from this normalization ambiguity,
but the experimental systematic error is dominated by it, and
our predictions for central \raa are not in disagreement with the data
within this large systematic uncertainty. 
Only a p + p run at $\sqrt{s}=2.76$ TeV at LHC will resolve this uncertainty.

In order to reduce 
systematic uncertainties, we therefore compared \rcp within the WHDG framework.  An \rcp analysis from ALICE was not released yet, and therefore
we calculated this quantity from the available results including 
propagating the reported uncertainties.  

The WHDG results for \rcp also show qualitative agreement with the \ptcomma-dependent shape of the data, 
although the WHDG calculations seem to somewhat under-predict the slope for the central and peripheral $\eqnraapt$ individually.  
On the other hand,
since  our central \raa calculations are more suppressed than the \raa
data while the peripheral results compare favorably, our \rcp results
also over-suppress the \rcp data.  This disagreement is at
about the 2-$\sigma$ level compared to the 15\% systematic error we
estimate  from the ALICE data.  Much of the 15\% systematic uncertainty is
due to the \ncoll normalization of \raa and is therefore
\ptcomma-independent and correlated, reducing the tension between the
experimental results and the theoretical predictions.  It is also
worth noting again that the \ncoll used by ALICE differs significantly from
that given by an optical Glauber calculation.
Our results would
compare more favorably with the data if 
the optical Glauber \ncoll
calculation would be used.  Precise experimental measurements of \highpt
probes that interact only weakly with the QGP plasma, such as direct
photons or Z bosons, is the only way of experimentally constraining
the \ncoll normalization used in \raa and \rcp analyses.  Z bosons
will give a much cleaner picture as photons are created not only
directly in the initial hard scatterings but also via photon
bremsstrahlung \cite{Zakharov:2004vm,Turbide:2005fk,Vitev:2008vk}.

As hard photons and Z bosons come from quark--anti-quark annihilation,
a measurement of $\eqnraa^\gamma$ or $\eqnraa^Z$ less than 1 could be
interpreted as due to either a reduction in the expected number of
binary collisions or the reduction in the availability of anti-quarks
due to gluon saturation.  In particular there are predictions of a
reduction by $\sim50\%$ in the initial hard spectra at the momenta
currently observed at LHC due to gluon saturation effects \cite{Albacete:2010bs}; this
reduction would naturally decrease as a function of \pt as $x$
increases.  As such, the \ptcomma-dependence of the ALICE data may be
due, at least in part, to initial state effects.  A precise
measurement of $\eqnraa^\gamma$ or $\eqnraa^Z$ showing the
\ptcomma-independence of these ratios would be required to demonstrate
the lack of strong initial state shadowing in the $x>0.01$ and $Q^2>100$ GeV$^2$ range relevant to the present ALICE data in the $p_T=10-20$ GeV range (as expected from, e.g., \cite{Paukkunen:2010qi}). p + A collisions would provide a superb independent test of the possible
influence of gluon saturation, similar to the crucial d + A collisions needed at RHIC to disentangle initial state from final state effects.  

Since there are no adjustable parameters for us, the
significant tension between our results and the ALICE data is a
failure to \emph{simultaneously} describe the normalizations of both
the RHIC and LHC \raaptcomma. One possibility is the sQGP produced at LHC is in fact more transparent than predicted by perturbative QCD tomographic models with medium densities that scale with observed particle rapidity densities.

Theoretical possibilities that could contribute to
the apparent transparency (decreased opacity) of the sQGP relative to the WHDG extrapolation from RHIC to LHC include
\begin{enumerate}
\item Baryon anomaly \cite{Vitev:2001zn,ToporPop:2007cg,ToporPop:2010qz}
\item Gluon feedback \cite{Vitev:2005he} 
\item Gluon to quark jet conversion \cite{Turbide:2005fk}
\item Gluon self energy \cite{Djordjevic:2003be,Djordjevic:2005nh}
\item Is the jet-medium coupling reduced at LHC:\\ $\alpha_s(LHC)< \alpha_s(RHIC)$ ?  
\end{enumerate}
Item 1 can be resolved when identified $\pi,K,p,\Lambda,\ldots,\Omega^-$
high-$p_T$ \raa becomes available. 
Item 2, in which the bremsstrahlung emitted gluons are kept track of and produce observed hadrons, may be a partial explanation,
but estimates so far have neglected the energy loss of the radiated gluons themselves. 
Radiated gluons with formation
length less than the size of the medium could also be strongly quenched.
A detailed centrality dependence of the di-hadron correlations 
may be able to resolve such nonlinear gluon shower feedback mechanisms.
Item 3 is possible channel that can effectively reduce 
the gluon jet color charge for asymmetric ($|x-0.5|\sim 0.5$) $q\bar{q}$ 
conversion. However, 
estimates\cite{Turbide:2005fk} so far 
have neglected strong interference effects of medium and vacuum 
induced amplitudes in finite size plasmas and high $p_T$ octet color coherence
of pair production. Jet flavor
triggers that could discriminate light quark and gluon jets 
are needed to determine experimentally if this mechanism
is responsible for the apparent transparency of the sQGP at LHC.
Item 4 involves HTL gluon dispersion effects on both the induced and hard
initial radiation associated with jet production. In  \cite{Djordjevic:2003be,Djordjevic:2005nh} it was shown that the Ter-Mikayelian and transition radiation effects reduce in general the effective dE/dL. Comparison of the centrality (or path length L) dependence
of \raa may help untangle such dispersion effects.

Item 5 refers to the possibility that the surprisingly transparent sQGP
at RHIC could be due a reduction of the effective jet-medium coupling 
between RHIC and LHC. In the WHDG analysis here,
we assumed $\alpha=0.3$ is the same at
RHIC and LHC and the \raa difference only reflects 
the increase of the initial sQGP density by 2.2. 
In \cite{Levai:2011qm} an average path length approximation using
GLV \cite{Gyulassy:2000fs,Gyulassy:2000er} 
was used to show that 
approximately the same effective opacity $L/\lambda\approx 6$ 
can fit both RHIC and LHC \raacomma. In the HTL approximation
$1/\lambda\propto T/\alpha^2$ up to slowly varying logarithmic corrections, so
at $L/\lambda$ constant implies $\alpha$ is reduced by a factor of $2.2^{1/3}=1.3$.
However, in  \cite{Levai:2011qm} $\alpha_s=0.3$ was assumed to be constant 
and $L$ varied. In WHDG the effective L is completely fixed the distribution of path lengths. As $\Delta E_{rad}\propto\alpha_s^3$ and $\Delta E_{el}\propto\alpha_s^2$ the theoretical calculation of suppression depends very strongly on the specific value of $\alpha_s$ taken. A fit, i.e.\ a postdiction to the LHC \raa data, can be achieved in WHDG by making the coupling a free parameter and reducing it from the $\alpha_s=0.3$ that we take in this analysis; we refrain from such uncontrolled fitting in this paper.  A quantitative estimate of the effect of allowing the coupling to run, and therefore possibly be smaller at LHC than at RHIC, requires higher order theoretical derivations that do not currently exist; in fact, even the qualitative result of such higher order effects are difficult to estimate as radiative energy loss calculations always include soft exchanges between the leading parton and the medium particles that, in principle can only be handled by nonperturbative techniques.  While one hopes that these higher order effects become small at higher leading parton momentum, there is always in heavy ion problems a temperature scale $T$ which is the same order of magnitude as $\Lambda_{QCD}$.  In particular, factorization has never been proven for energy loss calculations in heavy ion collisions.  Could jet coupling to the sQGP at LHC 
be in fact more weakly coupled than at RHIC?
From the near equality of bulk differentail elliptic flow, $v_2(p_T<2)$,
 the answer would appear to be no.
However, for short wavelengths $p_T>10$ GeV jet probes the effective jet medium 
coupling could in fact be smaller at LHC than at RHIC.
The key observable to test this possibility may be the high $p_T$ elliptic
and higher flow moments\cite{Wei:2009mj,Marquet:2009eq,Jia:2011pi,Betz:2011tu}, $v_2(p_T>10\;{\rm GeV/c})$. This observable remains a key stumbling block already at 
RHIC for all HTL/pQCD based models including WHDG that  underestimate $v_2(5\;{\rm GeV/c}<p_T<10\;{\rm GeV/c})$ significantly.
If $v_2(p_T>10\;{\rm GeV/c})$ at LHC turns out to deviate less from WHDG---even when
$\alpha(LHC)$ is reduced to account for the near identity of RHIC and LHC \raacomma---
then a firmer case could be made that the sQGP at LHC is indeed more transparent
to jets than expected.

In contrast to the above dynamical effects  that 
could contribute to an apparent reduction in opacity at LHC relative to our WHDG
expected growth $L/\lambda\propto (dN_{ch}/dy)^{1/3}$,  
there are other dynamical effects
neglected in our WHDG analysis that
could contribute to an apparent enhancement of the  sQGP opacity at LHC:
\begin{enumerate}
\item High $Q^2$ Color Glass Condensate \cite{Albacete:2010bs}
\item Dynamic magnetic scattering \cite{Djordjevic:2008iz,Djordjevic:2009cr,Buzzatti:2010ck}
\item AdS/CFT holography \cite{Gubser:2006bz,Herzog:2006gh,Chesler:2008uy,Gubser:2008as,Ficnar:2010rn}
\end{enumerate}
Item 1 can be constrained via dedicated p+A and A+A direct gamma and Z boson control  
experiments. Cross correlating light and heavy quark flavor jet quenching systematics would provide quantitative insight into the potential influence of Items 2 and 3. 

There are several other possible sources of uncertainty that we did
not address here.  For instance one might expect that the energy loss
in confined matter would reflect the different properties of a
hadronic medium as compared to a deconfined plasma of quarks and
gluons \cite{Domdey:2010id}.  Presumably, though, the cold matter energy
loss would be smaller than hot, and this would lead to a greater
discrepancy with the \rcp data.  Additionally, the ordering of length scales $1/\mu\ll \lambda_{mfp}\ll L$ assumed in the DGLV energy loss derivations is violated for short length paths that may contribute more substantially to the hadrons that are ultimately observed at LHC as compared to RHIC due to the significantly more dense medium.  Data from additional centrality classes will help clarify the possible role of final state confined matter effects and length scale ordering dependencies.  We also mentioned previously that
better calculations of the elastic energy loss of \highpt partons
exist and can additionally be improved on. 
Aside from the open physics
issues listed above, there is a continuing need to improve numerical
evaluation algorithms to remove simplifying numerical approximations
used in WHDG and other tomographic models. 
Due to the above considerations, experimental measurements of observables out to very high \ptcomma, for which we demand that theoretical calculations provide a consistent picture of both the mono- and di-jet data, will be crucial for furthering our understanding of the energy loss processes in experimentally accessible quark-gluon plasma, and hence crucially important for qualitatively and quantitatively determining the properties of these plasmas.

\section{Acknowledgments}
The work of MG was supported in part by DOE
Grant No.\ DE-FG02-93ER40764.  This work was supported by the U.S.\ DOE under Contract No.\ DE-AC02-05CH11231 and within the framework of the JET Collaboration \cite{JET}.

\def\eprinttmppp@#1arXiv:@{#1}
\providecommand{\arxivlink[1]}{\href{http://arxiv.org/abs/#1}{arXiv:#1}}
\def\eprinttmp@#1arXiv:#2 [#3]#4@{\ifthenelse{\equal{#3}{x}}{\ifthenelse{
\equal{#1}{}}{\arxivlink{\eprinttmppp@#2@}}{\arxivlink{#1}}}{\arxivlink{#2}
  [#3]}}
\providecommand{\eprintlink}[1]{\eprinttmp@#1arXiv: [x]@}
\renewcommand{\eprint}[1]{\eprintlink{#1}}
\providecommand{\adsurl}[1]{\href{#1}{ADS}}
\renewcommand{\bibinfo}[2]{\ifthenelse{\equal{#1}{isbn}}{\href{http://cosmolog%
ist.info/ISBN/#2}{#2}}{#2}}

\end{document}